\documentclass[journal]{IEEEtran}
\usepackage{cite}

 \usepackage{bbm}
\usepackage{color,soul}
\usepackage{gensymb}
\usepackage{dsfont}

\ifCLASSINFOpdf
  \usepackage[pdftex]{graphicx}
  \graphicspath{{../pdf/}{../jpeg/}}
  \DeclareGraphicsExtensions{.pdf,.jpeg,.png}
\else
  \usepackage[dvips]{graphicx}
  \graphicspath{{../eps/}}
  \DeclareGraphicsExtensions{.eps}
\fi
\usepackage[cmex10]{amsmath}
\usepackage{amssymb}

\usepackage{euscript}

\usepackage{multirow}
\usepackage{algorithm}
\usepackage[]{algpseudocode}
\usepackage{diagbox}
\usepackage{physics}
\usepackage{tikz}

\ifCLASSOPTIONcompsoc
  \usepackage[caption=false,font=normalsize,labelfont=sf,textfont=sf]{subfig}
\else
  \usepackage[caption=false,font=footnotesize]{subfig}
\fi
\usepackage{url}

\begin{document}
%
\title{Learning-Based Real-Time Event Identification Using Rich Real PMU Data}
%
%
%

\author{Yuxuan~Yuan,~\IEEEmembership{Student Member,~IEEE,}
    Yifei~Guo,~\IEEEmembership{Member,~IEEE,}
	Kaveh~Dehghanpour,~\IEEEmembership{Member,~IEEE,}
	Zhaoyu~Wang,~\IEEEmembership{Member,~IEEE,}
	and Yanchao~Wang
\thanks{This work is supported by the U.S. Department of Energy Office of Electricity under DEOE0000910. (\textit{Corresponding author: Zhaoyu Wang})

 Y. Yuan, Y. Guo, K. Dehghanpour, Z. Wang, and Y. Wang are with the Department of
Electrical and Computer Engineering, Iowa State University, Ames,
IA 50011 USA (e-mail: yuanyx@iastate.edu; wzy@iastate.edu).
 }
}
%
%

\markboth{Submitted to IEEE for possible publication. Copyright may be transferred without notice}%
{Shell \MakeLowercase{\textit{et al.}}: Bare Demo of IEEEtran.cls for Journals}
%



\maketitle

\begin{abstract}
A large-scale deployment of phasor measurement units (PMUs) that reveal the inherent physical laws of power systems from a \emph{data} perspective enables an enhanced awareness of power system operation. However, the high-granularity and non-stationary nature of PMU time series and imperfect data quality could bring great technical challenges to real-time system event identification. To address these issues, this paper proposes a two-stage learning-based framework. At the first stage, a Markov transition field (MTF) algorithm is exploited to extract the latent data features by encoding temporal dependency and transition statistics of PMU data in graphs. Then, a spatial pyramid pooling (SPP)-aided convolutional neural network (CNN) is established to efficiently and accurately identify operation events. The proposed method fully builds on and is also tested on a large real dataset from several tens of PMU sources (and the corresponding event logs), located across the U.S., with a time span of two consecutive years. The numerical results validate that our method has high identification accuracy while showing good robustness against poor data quality.
\end{abstract}


\section{Introduction}\label{introduction}
Large-scale blackouts, such as the Northeast blackout of 2003 in the U.S., which started with a local fault but eventually affected 50 million customers, continuously remind us of the need for better and faster event detection and identification to enhance the wide-area situational awareness of power system operation \cite{2003bo}. Recent years have seen a rapid growth in the deployment of phasor measurement units (PMUs), providing a unique opportunity for preventing cascading failures and blackouts \cite{Ian}. Unlike the supervisory control and data acquisition (SCADA) system that only offers power system monitoring at steady state, PMU collects the precisely synchronized voltage and current phasor, frequency, and frequency variation at sub-second intervals (e.g., 30 or 60 samples per second in U.S.), which enables capturing the fast dynamics of power systems. Therefore, exploiting PMU data for real-time event identification has attracted an increasing attention.

\emph{Related Works:} The existing works on PMU-based event \emph{detection} and \emph{identification} can be mainly classified into two categories: 1) traditional signal processing-based methods \cite{DK2017, MC2019, YG2015, EP2008}; and 2) machine learning-based methods \cite{MB2016,SL2020,JM2012,SB2017}. In \cite{DK2017}, a wavelet-based method was designed for detecting the event occurrence and classifying events. In \cite{MC2019}, a dynamic programming-based swinging door trending method was developed to detect the start-time and placement of events. The authors in \cite{YG2015} proposed a quadratic fitting method to recover the  dynamics of events and a knowledge-based criterion to classify events. In \cite{EP2008}, the extended Kalman-filtering algorithm was applied to detect voltage events. Inspired by the recent success of machine learning techniques in data analytics, many researchers have adopted different machine learning methods to identify the types of events. In \cite{MB2016}, a multiclass extreme learning machine classifier was utilized to perform near-real-time automatic event diagnosis. In \cite{SL2020}, a data-driven algorithm consisting of an unequal-interval reduction method and principal component analysis was proposed to detect and locate events using PMU data. In \cite{JM2012}, a hierarchical clustering-based method was proposed to determine the types of events, using several characteristics of multidimensional minimum volume enclosing. In \cite{SB2017}, the k-nearest neighbor and support vector machine classifiers were exploited to perform event identification based on different pattern creation methods. 

\emph{Challenges:} While researchers have contributed numerous valuable works on this topic, several critical questions remain open, which might challenge the practical deployment of the methods. 1) Most of the signal processing-based methods require massive computations due to the complicated mathematical transformation and optimization, which may limit their real-time applications \cite{SL2020}. 2) Machine learning-based methods typically suffer from event data scarcity, resulting in a \textit{data imbalance} problem \cite{LH2019}. This could hinder reliable training of learning-based models \cite{VC2007}. 3) Due to the fact that real PMU data and the event log counterpart might not be available, most of the previous studies relied on simulated data, which may not be able to precisely replicate the real cases. This challenges the real-world deployment of such methods. 4) The impact of PMU data quality problems caused by unsuccessful communications (e.g. missing data, bad data, and time error) on event identification process was not addressed in previous works. For example, data missing can disjoint the dimensional consistency of data samples between the offline training and online testing. As illustrated in our later analysis of large amount of real PMU recordings, such problems are inevitable and not rare. Poor robustness against data quality makes the event identification models not sufficiently convinced \cite{JZ2019}.

\begin{figure}[tbp]
	\centering
	\includegraphics[width=3.5in]{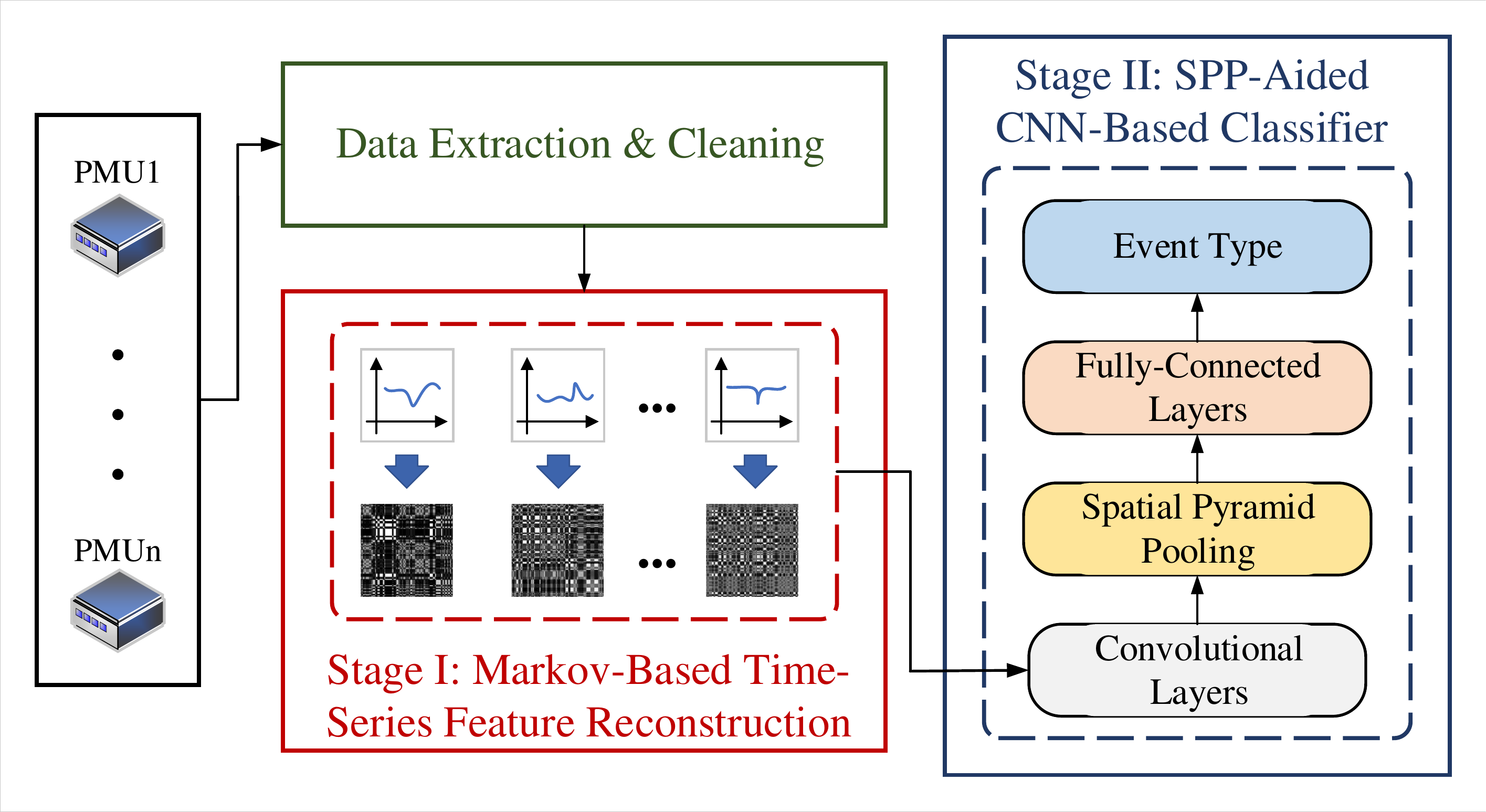}
	\caption{Illustration of two-stage event identification framework. In the data extraction and cleaning, a 2 s time window is selected to extract the event data and then PMU status information and engineering intuition are utilized to eliminate the missing and/or bad data. The stage I encodes the PMU data to a graph by characterizing the transition probability and temporal dependency. The stage II constructs an end-to-end mapping between the graphs and the event types by leveraging CNNs.}
	\label{fig:main}
\end{figure}

\emph{Our Contributions:} To develop a convinced event identifier that does not suffer from the weaknesses led by data scarcity or simulation-based modeling, we build the model fully by a large amount of real-world (voltage and frequency) data over two consecutive years, gathered from several tens of PMUs throughout the U.S.. So, such data enjoys both the richness of \emph{length} and \emph{width}, which implies there are not only adequate event samples (two years) but also redundant features (multiple PMUs). This will significantly improve the reliability of the classifier. However, numerous data will, in turn, bring many technical challenges to the data quality and algorithm efficiency. To address these challenges, we propose a two-stage PMU-based event identification framework (see Fig. \ref{fig:main}): 1) Markov-based time-series feature reconstruction; and 2) spatial pyramid pooling (SPP)-aided convolutional neural network (CNN)-based event classification. The first-stage Markov-based feature reconstruction is able to capture the \emph{time-varying} statistical characteristics of PMU time series and should be inherently robust to data noise thanks to its probabilistic nature. In the second stage, the parameter sharing and pooling of the CNN-based classifier significantly reduce the number of learning parameters, thus, ensuring the algorithm efficiency and scalability. The last but not the least, the proposed method is capable of allowing the signals of arbitrary dimensions during online testing, which is of great importance for real-life implementations. To the best of our knowledge, this is the first attempt to develop an efficient event identification method by leveraging such large-scale real PMU dataset and event logs\footnote{It is worth noting that a few remarkable works, e.g., \cite{DK2017} and \cite{SB2017}, developed the learning models using real PMU data. E.g., the authors in \cite{SB2017} conducted the analyses using the real PMU dataset from Public Service Company of New Mexico, which, however, only contains 84 event labels recorded by four PMUs. From the perspective of machine learning, such data amount might not be sufficient for developing a stable classifier in practice.}. The numerical results demonstrate the classification accuracy of the proposed model. We also test the robustness of our method to online data quality problems.


The rest of this paper is constructed as follows: Section \ref{overall} introduces the available PMU dataset and data pre-processing. In Section \ref{MTF}, an Markov-based time-series feature reconstruction algorithm 
is utilized to summarize the hidden features of PMU data in graphs. Section \ref{SPPCNN} proposes the SPP-aided CNN-based event identification method based on MTF-graphs. The numerical results are analyzed in Section \ref{result}. Section \ref{conclusion} presents research conclusions.

\section{Data Description and Pre-Processing}\label{overall}
\subsection{PMU Dataset Description}
The available PMU dataset includes more than 440 PMU sources that are installed in the Eastern, Western, and Electric Reliability Council of Texas interconnections at different voltage levels, i.e., 69, 115, 138, 161, 230, 345, 500 and 765 kV, with the nominal frequency of 60 Hz. For convenience, let A, B and C denote the three interconnections hereinafter. They are equipped with 215, 43 and 188 PMUs, respectively. Most data segment is archived at 30 frames/s and the remaining is archived at 60 frames/s. Each PMU measures voltage and current phasor, system frequency, frequency variation rate and PMU status information. The dataset spans a time period of around two consecutive years (2016--2017). The total size of the dataset is more than 20 TB (in Parquet form)\footnote{The research team has processed the raw dataset to 20 TB in Parquet form so as to save memory while facilitating the learning algorithm design and test.}. These data files were read in Python and MATLAB environments. In total, around 670 billion sampling points will be used to conduct the analyses. 
\subsection{Event Log Description}
 Event logs play a vital role in providing the ground truths for investigating event identification methods. However, event labels are scarce and are generally not openly available for most cases, which truly becomes a big barrier for research purpose. Total 6,767 event labels, including 6,133 known events and 634 unknown events (where the event type entry is empty), are imported and collected from SCADA systems. Besides, the types and timestamps of the events have been verified by matching them with the corresponding protection relay records, ensuring the high confidence of the event logs.\footnote{The event logs (labels) are provided by several transmission organizations (TOs) and verified by system operators.} Each event log contains the interconnection number, start timestamp, end timestamp, event type, as well as event cause, of which a detailed statistical summary is presented in Table \ref{table:1.1}. It should be noted that the three interconnections have different event categorization systems. Hence, the datasets from these interconnections cannot be combined into one dataset. By assessment, the PMU data from interconnection B shows the best data quality and therefore, will be utilized to develop and validate our method in this paper.

\subsection{Data Pre-Processing}
Data quality issues such as bad data, dropouts, communication issues, time error, can lead to misclassification of bad data as events, thus, decreasing the accuracy of the method. Hence, the PMU dataset is initially passed through data pre-processing. The goal of the data pre-processing is twofold: 1) select an analysis-window to extract the data into frames corresponding to pre-event and event states for training a learning model; 2) eliminate missing and bad data caused by communication and meter malfunction.
\begin{table}[tbp]
\caption{Statistical Summary of Dataset From 440 PMU Sources.}
\centering\renewcommand\arraystretch{0.75}
\setlength{\tabcolsep}{0.2mm}{
\begin{tabular}{p{4cm}p{1.2cm}p{1.2cm}p{1.2cm}}
\hline\hline\\[-5pt]
 & A & B & C\\[0pt]
\hline\\[-5pt]
Record period & 1 year & 2 years & 2 years\\[2pt]
Data size & 3 TB & 5 TB & 12 TB\\[2pt]
Number of PMUs & 215 & 43 & 188\\[2pt]
Sample rates [frames/s] & 30  & 30/60 & 30 \\[2pt]
Total number of events & 29 & 4854 & 1884\\[2pt]
Number of unidentified events & 0 & 0 & 634\\[2pt]
Resolution of event record & Daily & Minute & Minute\\[2pt]
Number of event causes & 13 & 3911 & 1883\\[2pt]
\hline\hline
\end{tabular}}
\label{table:1.1}
\end{table}
\begin{figure}[tbp]
\centering
\subfloat [Frequency event example.]{
\includegraphics[width=3.3in]{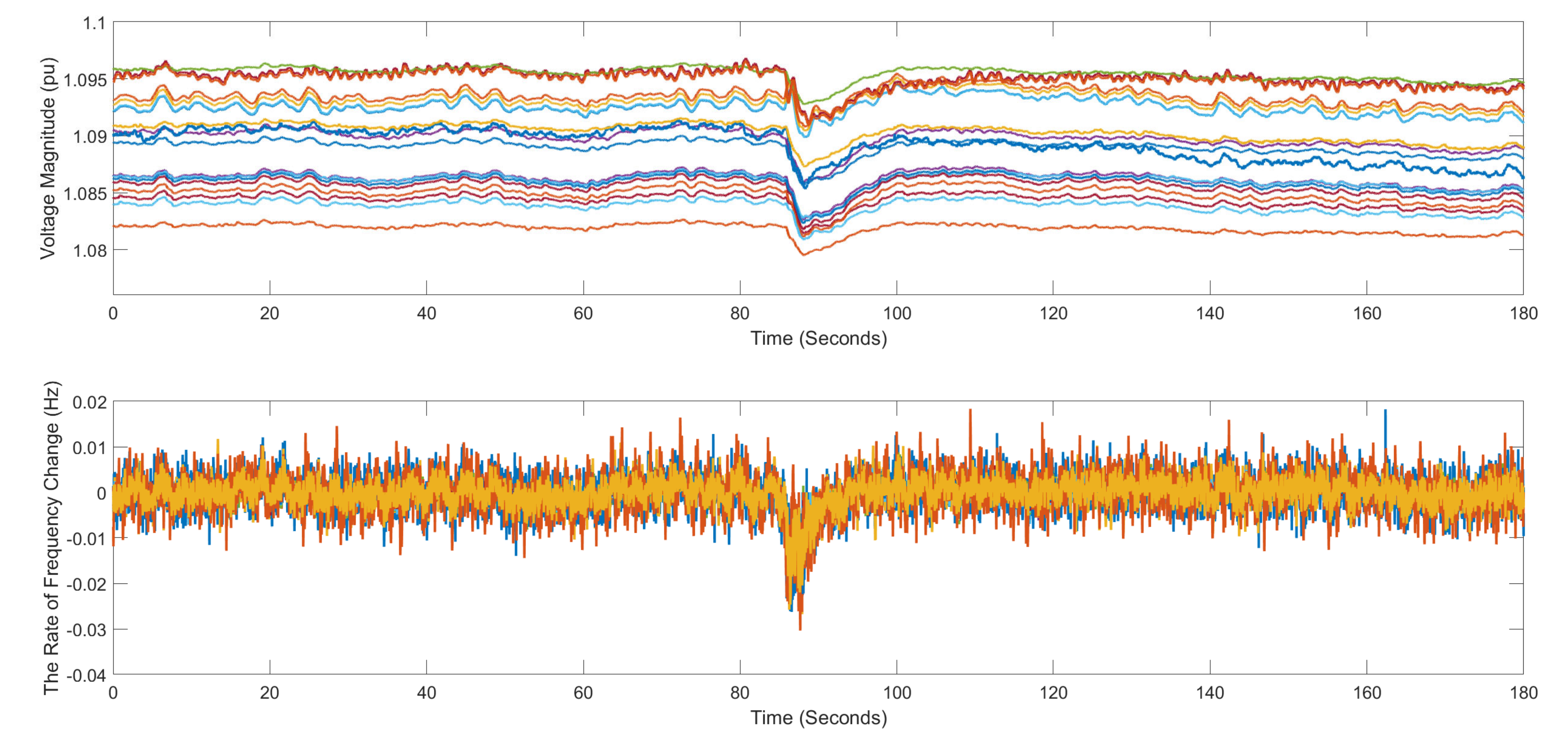}
}
\hfill
\subfloat [Line outage example.]{
\includegraphics[width=3.3in]{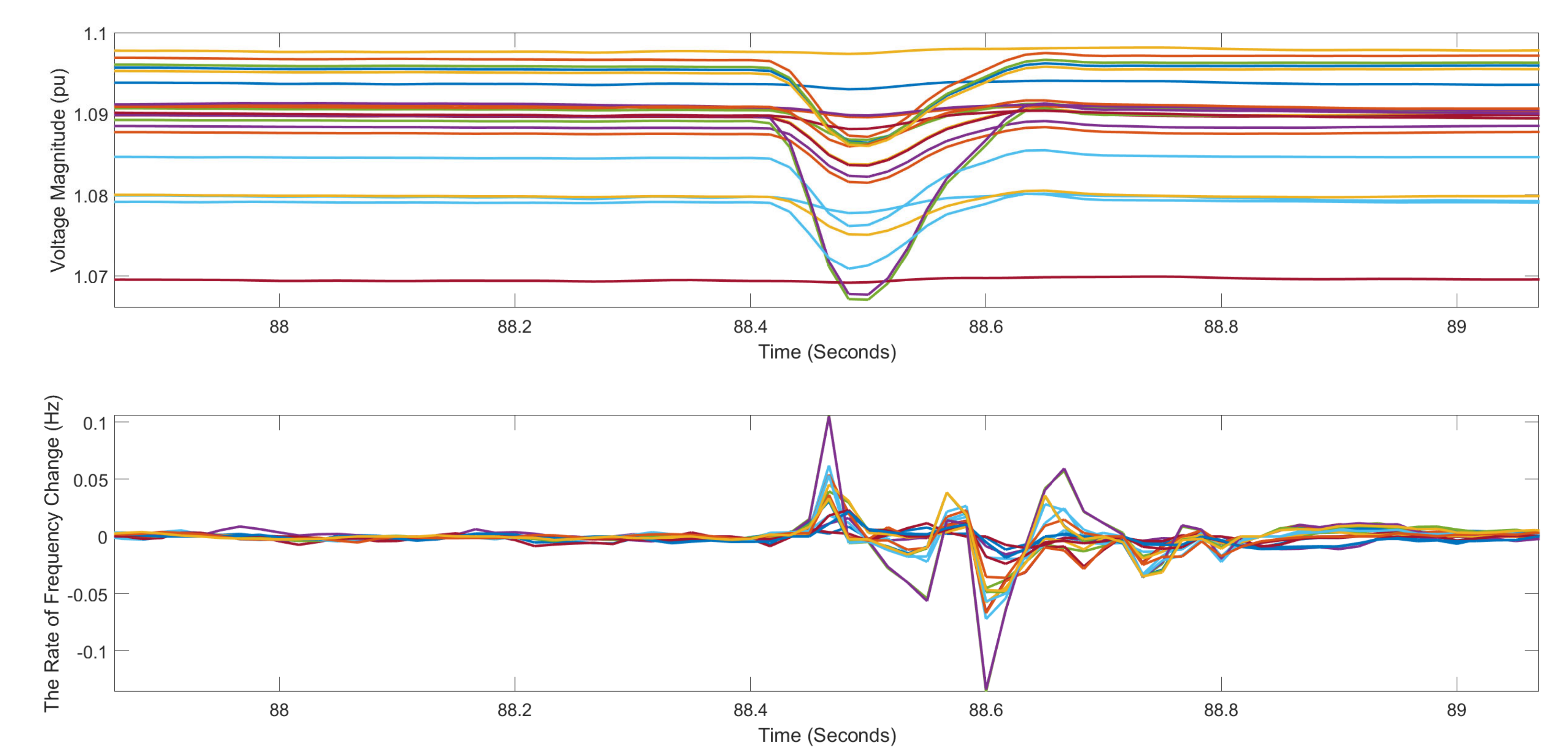}
}
\caption{Plots of multiple PMU data for different events.}
\label{fig:event_example}
\end{figure}
\begin{figure*}[tbp]
      \centering
      \includegraphics[width=2\columnwidth]{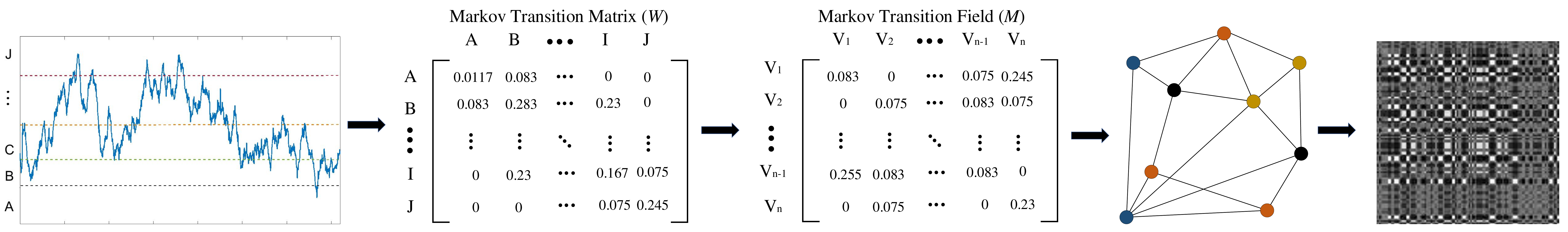}
\caption{Illustration of the proposed encoding map of MTF. As shown in Fig. \ref{fig:MTF}, the square matrix $M$ can be interpreted as a network $\mathcal{G}$, where $m_{k_1,k_2}$ represents weight of the edge between any two vertices $k_1$ and $k_2$. The nodes in different colors precisely match different time points of $V^j_i$. }
\label{fig:MTF}
\end{figure*}

Following the start timestamps in event logs, we have extracted 60 seconds of pre-event and 120 seconds of post-event data to visualize event pattern. Note that the frequency variation and voltage magnitude data are utilized as indicators of power events based on previous research \cite{DK2017}. Fig. \ref{fig:event_example} shows event plots of all PMUs in the system. As is demonstrated in Fig. \ref{fig:event_example}, it is clear that the most critical changes happen around the inception of event, but the lengths of changes are different for different PMU-recordings. In addition, these figures show that the lengths of changes can be at second- or sub-second-levels for different types of events. Thus, to apply PMU-base event identifiers in real-world application, a second-level of analysis-window is needed. In this work, a 2-second analysis-window is selected to extract the event data \cite{SB2017,OPD2012}. According to the sampling rate of PMUs, each event window includes 120 data points. However, the resolution of available event logs is minute-level, thus, not sufficient for directly extracting the start timestamps of events at the second-level. To tackle this, a statistical algorithm is proposed that detects the transition between the normal and event states. The rationale behind this is that, since PMUs are synchronized, the variations in PMU-recordings will occur at the same time. It should be noted that this statistical algorithm can be bypassed if the resolution of event logs is sufficient for a 2-second analysis-window. The proposed algorithm involves the following steps:
\begin{itemize}
\item \textbf{Step 1:} Define and initialize the 2-second event set $\mathbbm{E} = \emptyset$ and the event counter $i\gets 1$.
\item \textbf{Step 2:} Select the $i$'th event from the event logs and then extract related 60 seconds of pre-event and 120 seconds of post-event data $\mathbbm{D}_i$.
\item\textbf{Step 3:} Utilize the modified $z$-score for $\mathbbm{D}_i$ and identify the time stamps with the minimum score, of which the set is denoted as $\mathbbm{T}_i$ \cite{sat2006}.
\item \textbf{Step 4:} Find the time stamp with the highest frequency of minimum values belonging to $\mathbbm{T}_i$, denoted as $t^\ast_i$.
\item \textbf{Step 5:} Sort $\mathbbm{D}_i$ based on the 2-second analysis-window, and find the 2-second data that includes $t^\ast_i$, denoted as $\mathbbm{D}^\ast_i$; add $\mathbbm{D}^\ast_i$ to $\mathbbm{E}$.
\item \textbf{Step 6:} $i \gets i+1$; go back to Step 2 until $i$ equals the total number of events.
\end{itemize}

When the 2-second event dataset is obtained, PMU status flags information and engineering intuition are utilized to perform data cleaning. The status flags are in binary form and all information is aligned as 16-bit long. Each bit corresponds to a different status based on IEEE C37.118.2-2011 standard, such as bits 03-00 reflecting the trigger reason and bits 05-04 showing the time error (i.e., asynchrony). When the value of the status flag is 0 in the decimal format, data can be used properly; otherwise, data should be removed due to the various PMU malfunction. Also, engineering intuition provides additional insights into the data quality issues which are not identified by the status flags. For example, a number of data windows contained a single sample with an unreasonable value compared to the nominal value, which was dismissed as bad data. In this work, when the total number of missing/bad data is found to be greater than $5\%$ of the total number of data samples, the event is excluded from our study. Furthermore, the rest of the missing/bad data are filled and corrected through interpolation \cite{OPD2012}. 

\begin{figure}[tbp]
\centering
\subfloat [Survival function of probability of missing/bad data]{
\includegraphics[width=3in]{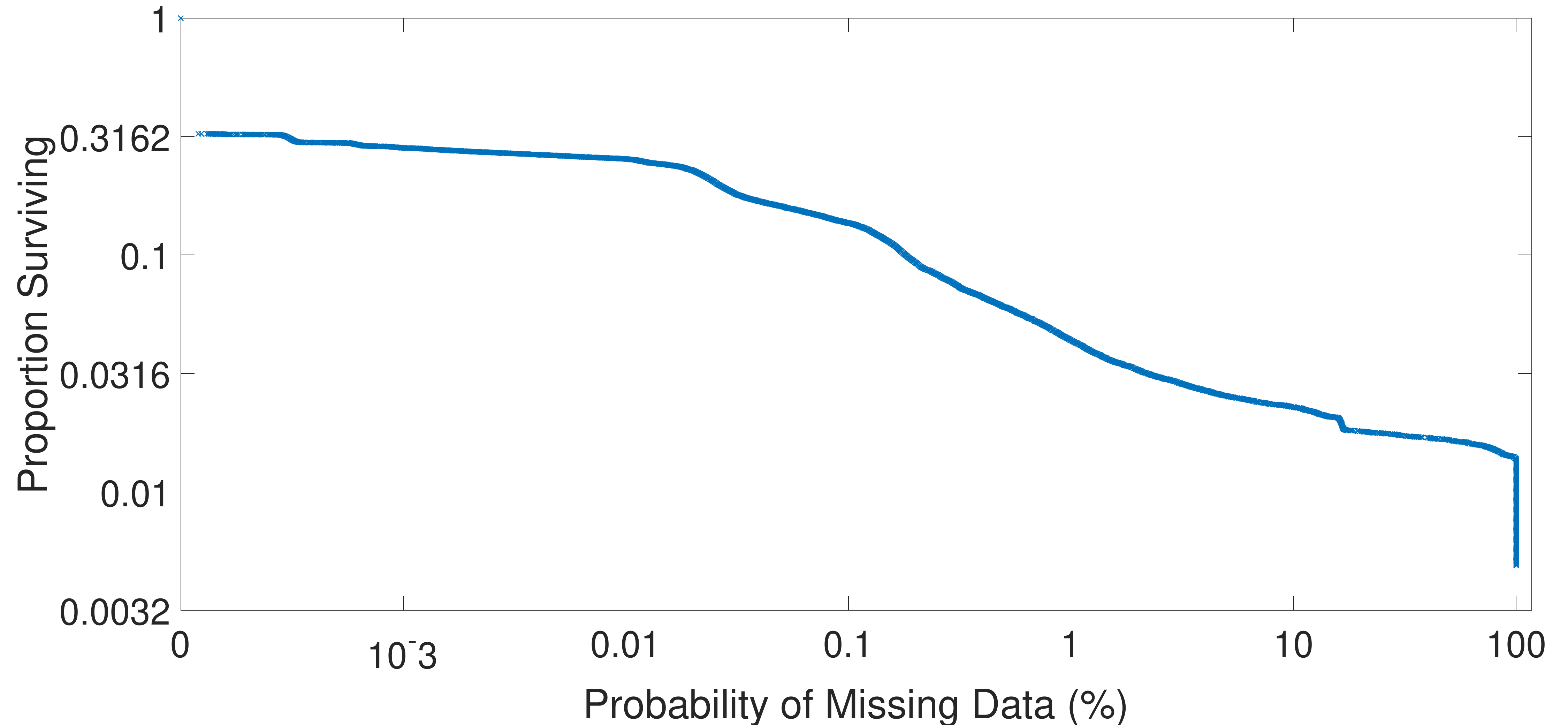}
}
\hfill
\subfloat [Survival function of size of single data quality problem]{
\includegraphics[width=3in]{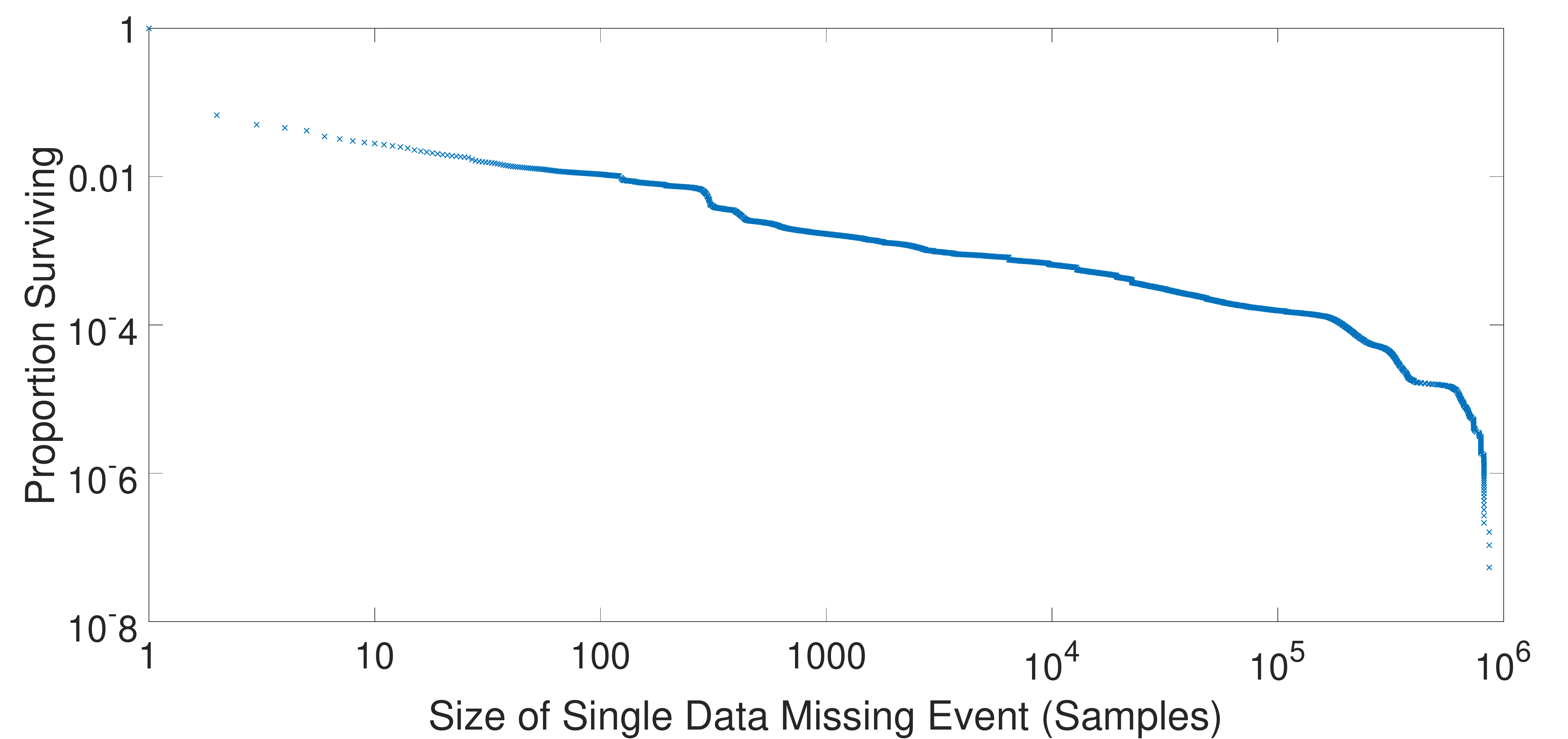}
}
\caption{Statistical results for data quality problem.}
\label{fig:stat}
\end{figure}

\section{Markov-Based Time-Series Feature Reconstruction}\label{MTF}
Despite PMUs' high precision and ability to capture system dynamics, PMU-based event identification via simple features (i.e., voltage magnitude and frequency) is a difficult task. The source of this challenge is the non-stationary characteristics of real-world PMU data, which is caused by sudden variations in system behavior during events \cite{DK2017}. To address this issue, different signal processing techniques are utilized to perform feature reformulation, such as discrete Fourier transform, discrete wavelet transform, and multidimensional minimum volume enclosing ellipsoid \cite{SB2017}. In this paper, a Markov matrix-based method known as MTF is adopted to encode the temporal dependency and transition statistics of PMU data in a compact metric \cite{WZ2015}.

Let $V^j_i$ denote the voltage magnitude data during event $i$ as recorded by the $j$'th PMU. The objective of the proposed feature reformulation method is to map this continuous signal $V^j_i=\{V^j_i(k)|k\in \mathbbm{N}, V^j_i(k) \in \mathbbm{R}\}$ to a network $\mathbbm{G} = (\mathbbm{O},\mathbbm{B})$, which consists of a set of vertices $\mathbbm{O}$ and a set of edges $\mathbbm{B}$ connecting different vertices. Since a direct mapping from continuous data to a network with finite nodes is impossible, we utilize a quantile-based approach to obtain a discretized dictionary for $V^j_i$ \cite{CA2011}. Specifically, given a $V^j_i$, we create $q$ quantile bins (states) $S_1,...,S_q$ and assign each $V^j_i(k), k=1,...,n$, to the corresponding bins\footnote{Note that, $S_1,...,S_q$ are different for different $i,j$. For simplicity, we omit the indexes $i,j$ here.} (see Fig. \ref{fig:MTF}). While different strategies can be applied to assign $V^j_i$ to the bins, our quantile strategy ensures that all bins in each data have the same number of points \cite{CA2011}. Compared to other strategies, quantile mapping is more data-specific and has shown the highest identification accuracy on our dataset. Following this strategy, a weighted adjacency matrix $W \in\mathbbm{R}^{q\times q}$ is developed by counting the transitions among quantile bins similar to a first-order Markov chain. Each entry of $W$ is a non-negative real number representing a transition probability that is determined as follows:
\begin{align}
\nonumber w_{S_a,S_b}=&{\rm Pr}\left\{V^j_i(t)\in S_a|V^j_i(t-1)\in S_b\right\},\\
&\hspace{10mm}\forall S_a\in\{S_1,...,S_q\},S_b\in\{S_1,...,S_q\}.
\end{align}
After normalization by $\sum_{S_b}{w_{S_a,S_b}}=1$, $W$ becomes a standard Markov matrix that contains the transition probability on the voltage magnitude axis. However, $W$ fails to capture the higher order temporal dependencies as it is based on a first-order Markov chain. Hence, to preserve information across the temporal dimension, we extend matrix $W$ to a new matrix $M\in \mathbbm{R}^{n\times n}$ by aligning each probability along the temporal order, as follows \cite{WZ2015}:
\begin{align}
M=
\begin{bmatrix}
m_{11} & \cdots & m_{1n}\\
\vdots & \ddots & \vdots\\
m_{n1} & \cdots & m_{nn}
\end{bmatrix}     
\end{align}
with 
\begin{align}
    \nonumber m_{k_1,k_2}=w_{S_a,S_b},\,\,V^j_i(k_1)\in S_a, V^j_i(k_2)\in S_b, \forall k_1,k_2.
\end{align}
So, the $k$th row of $M$ represents the transition probabilities between the $k$th point and all data points. In this way, $M$ encodes the transition dynamics of the PMU data between different time lags. This process is applied to the remainder of event dataset including voltage magnitudes and frequency variations to obtain the MTF-based graph set, which are used for training a learning model.

\begin{figure}[tbp]
	\centering
	\includegraphics[width=3.2in]{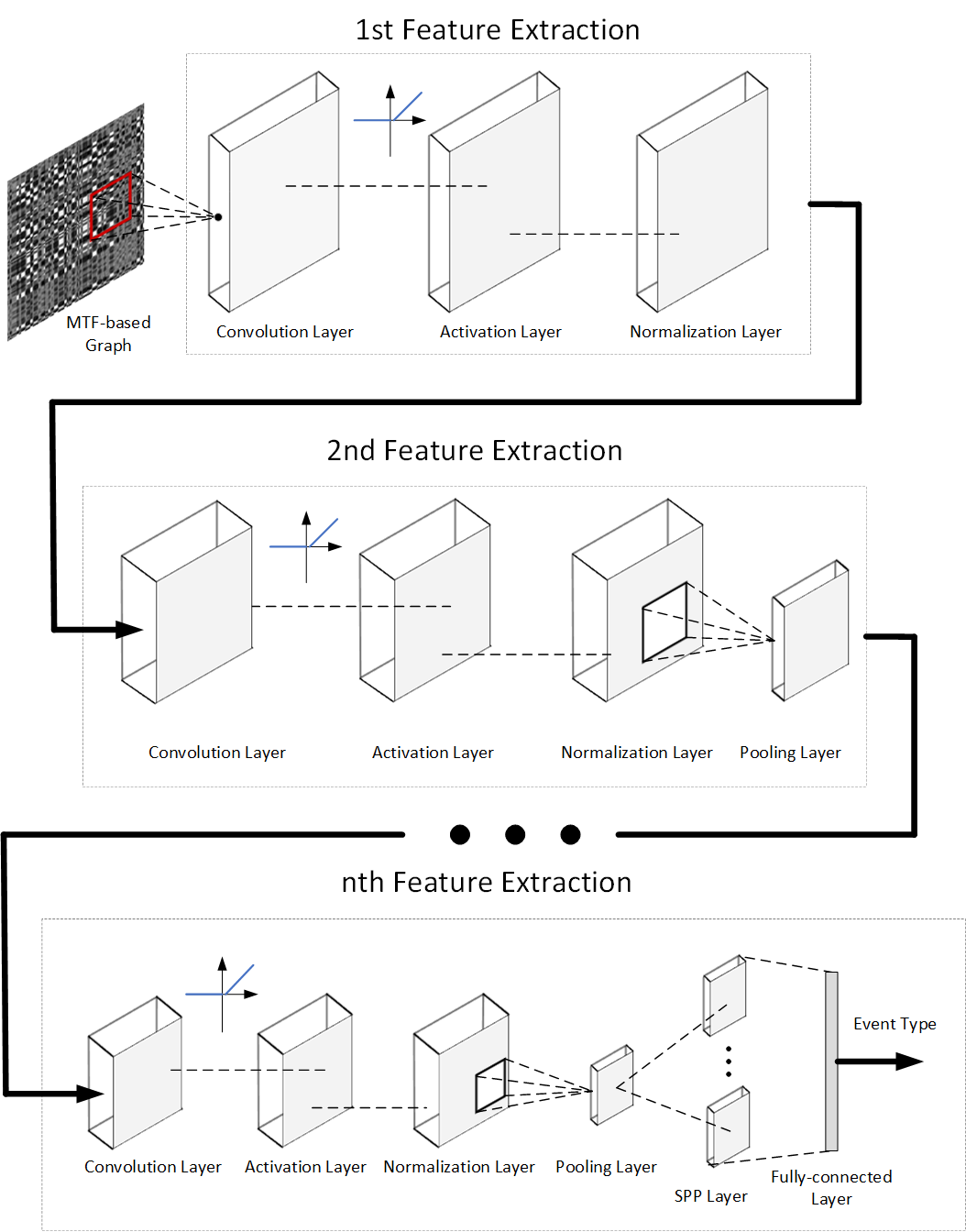}
	\caption{Proposed SPP-aided CNN-based event classifier. As can be seen, our model is a multiple-layer architecture that consists of different layers. The input of this mode is the MTF-based graphs and the outcome is the event type.}
	\label{fig:spp_cnn}
		\vspace{-1em}
\end{figure}

\section{SPP-Aided CNN-Based Event Identifier}\label{SPPCNN}
In this section, we lay out our PMU-based event identification strategy. Considering that PMU-based models are developed to identify events and perform supervisory protection in real-time, high speed and accuracy are required \cite{SB2017}. Also, in practice, data quality problems should be considered in algorithm design. In order to analyze PMU data quality problems, we leverage a \textit{survival function}, $S(k)$. This function is defined for the probability of missing data per PMU per day as follows:
\begin{equation}
\label{eq:sur1}
S(k) = {\rm Pr}\left\{\frac{\mbox{number of missing data per PMU per day}}{\mbox{total number of data per PMU per day}} > k\right\}.
\end{equation}
The empirical survival function computed directly from our PMU dataset is shown in Fig. \ref{fig:stat} (a). As can be seen in this figure, PMUs show data quality issues more than 30$\%$ of time. Similarly, the survival function of size of each individual data quality problem is plotted in Fig. \ref{fig:stat} (b). Based on this figure, around $3\%$ of data quality issues have more than 10 consecutive missing data. Considering the extremely high sampling rate of PMU, it is quiet common to have consecutive missing data due to long communication failure intervals or equipment malfunction. When a data quality problem occurs, the corresponding PMU recordings are seen as ``NaN" in the data file. For offline training, the PMU signals that contain the ``NaN" can be removed to mitigate the data quality problems. Nonetheless, during online testing, most data-driven models cannot be applied since these models only accept inputs with fixed dimension. In other words, the testing input dimension of the models should be equal to that of the training data (i.e., if $n$-dimensional data is used for training, then the data-driven model allows for $n$-dimensional test inputs.). One common solution is to perform data imputation to fill the ``NaN" values in order to keep the input dimension fixed. However, no matter how advanced these data imputation methods are, they typically generate additional inference errors, which reduce the accuracy of event identification. Thus, efficient handling of data quality problems is of vital importance for real-time event identification. To tackle this, an SPP-aided CNN-based event classification framework is proposed that utilizes the recently-developed deep learning techniques, as shown in Fig. \ref{fig:spp_cnn}. Our method can construct an end-to-end mapping relationship between MTF-based graphs and the event types. Moreover, this method introduces robustness to data quality problems during online testing.
\begin{figure}[tbp]
	\centering
	\includegraphics[width=3.5in]{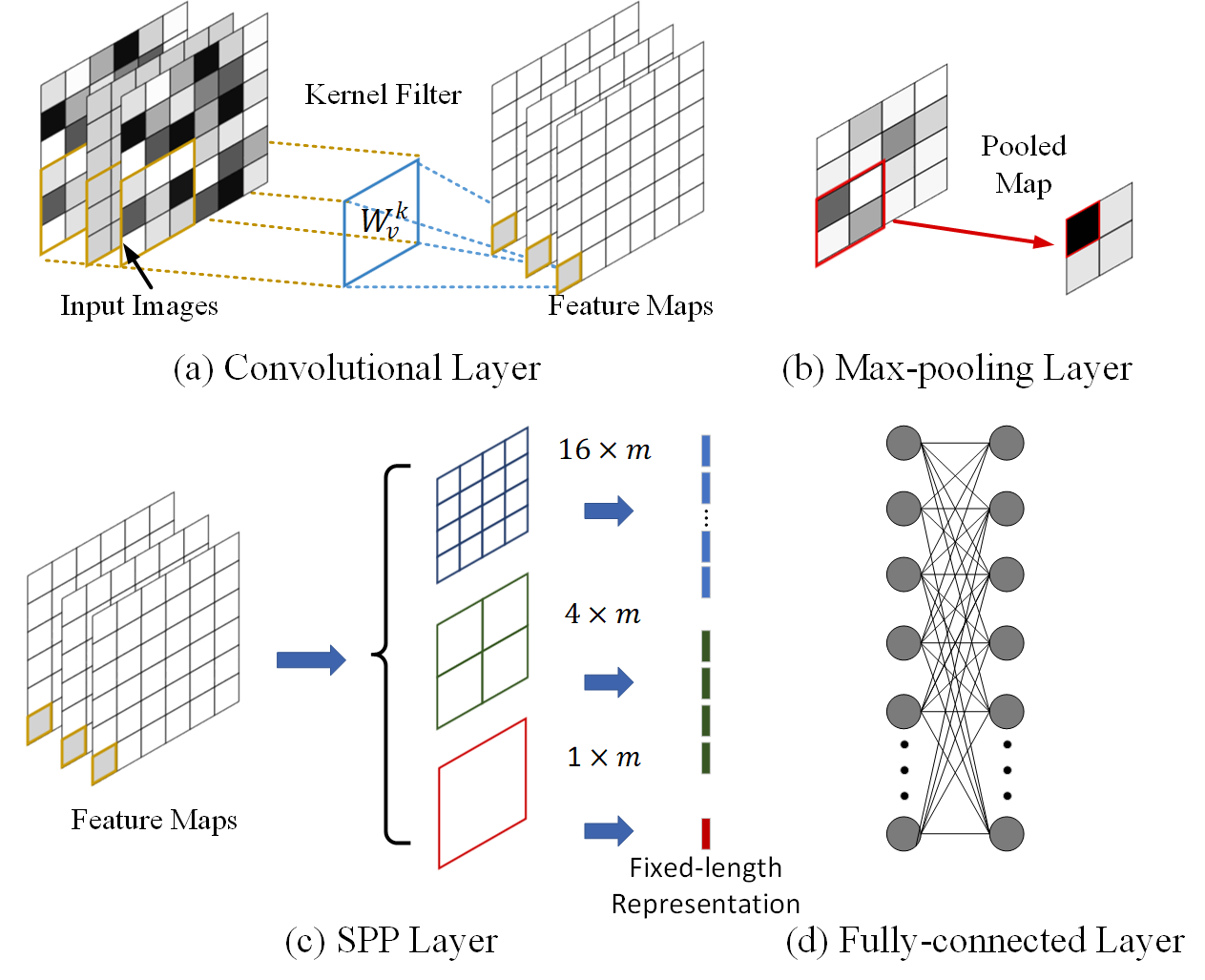}
	\caption{Illustrate of the different layers in the proposed model; (a) Convolutional Layer; (b) Max-Pooling Layer; (c) SPP Layer (d) Fully-Connected Layer.}
	\label{fig:layer}
\end{figure}

Here, consider a training set $\{\mathbbm{V},\mathbbm{F},\mathbbm{L}\}$ where $\mathbbm{V}:=\{v^{(1)},...,v^{(h)}\}$ and $\mathbbm{F}:=\{f^{(1)},...,f^{(h)}\}$ are the MTF-based graphs using PMU-based voltage magnitude and frequency variation data, and $\mathbbm{L}:=\{l^{(1)},...,l^{(h)}\}$ is the corresponding event label set from the event logs. Then, the probability that the label $l^{(i)}$ of $\{v^{(i)},f^{(i)}\}$ is equal to $j$ can be calculated by:
\begin{equation}
\label{eq:pro}
{\rm Pr}\left\{l^{(i)}=j|z^{(j)}\right\}=\frac{{\rm exp}\left({\theta_j(v^{(i)},f^{(i)})}\right)}{\sum_{c=1}^o{\rm exp}({\theta_c(v^{(i)},f^{(i)})}}
\end{equation}
where, $o$ is the number of event types and $\theta_c(\cdot)$ denotes the mathematical model in the proposed SPP-aided CNN method. The learning parameters by minimizing the following cost function $J$:
\begin{equation}
\label{eq:cost}
J:= -\frac{1}{h}\sum_{i=1}^h\sum_{j=1}^o\mathbbm{1}\{j=l^{(i)}\}\mbox{ln}\left(\frac{{\rm exp}({\theta_j(v^{(i)},f^{(i)}))}}{\sum_{c=1}^o{\rm exp}({\theta_c(v^{(i)},f^{(i)}))}}\right)
\end{equation}
where $\mathbbm{1}\{j=l^{(i)}\}$ equals 1, if $j$ equals $l^{(0)}$; otherwise, it is 0. Here, $\theta(\cdot)$ consists of multiple convolutional, batch normalization, max-pooling, SPP, and the fully-connected layers. 

\textbf{Convolutional Layer:}
The key component of the convolutional layer is the convolution operation: $\ast$. This layer computes convolutions of the input with a series of filters, which can be mathematically described as follows \cite{auto2017}:
\begin{equation}
\label{eq:convof}
\phi_{g}^\zeta = \sum_{u \in U}x_{g-1}^u\ast W_{g}^\zeta+b_{g}^\zeta
\end{equation}
where $\phi_{g}^\zeta$ is the latent representation of the $\zeta$th feature map of the $g$th layer (the first feature map is the input graph $\{v^{(i)},f^{(i)}\}$); $x_{g-1}^u$ is the $u$th feature map of the previous layer and $U$ is the total number of feature maps; $W_{g}^\zeta$ and $b_{g}^\zeta$ are the kernel filter and the bias of the $\zeta$th feature map of the $g$th layer, respectively. Since all event signals have been regarded as 2-dimensional MTF-based graphs after the feature reformulation, $x_{g-1}^u\ast W_{g}^\zeta$ can be written as \cite{Goodfellow2016},
\begin{equation}
\label{eq:convo}
(x_{g-1}^u\ast W_{g}^\zeta)(i,j)=\sum_{\delta_{i}=0}^{U-1}\sum_{\delta_{j}=0}^{U-1}x_{g-1}^s(i-\delta_{i},j-\delta_{j})W_{g}^\zeta(i,j) 
\end{equation}
where $i$ and $j$ are the row and column indices of the MTF-based graphs. Hence, the convolutional layer operates in a sliding-window way to output the feature maps (see Fig. \ref{fig:layer}(a)) \cite{SPP}. The amount of horizontal and vertical movement in the sliding-window is set to 1 here. For each convolutional layer, the size of the output feature map is  $\phi_{g}^\zeta \in \mathbbm{R}^{(p-q+1) \times (p-q+1)}$ where $x_{g-1}^u$ and $W_{g}^\zeta$ are $p \times p$ and $q \times q$ matrices, respectively. A typical drawback of the convolutional layer is that the impact of the data samples located on the border of data graph is much smaller than those at the center. To tackle this, a \textit{padding strategy} is used by adding an additional layer to the border of the feature maps \cite{YQ2016}. 

\textbf{Activation Layer:}
To make up for the limitation of linear modeling in the convolutional layer, the outcomes of $g$'th convolutional layer are passed to an activation layer. A nonlinear function, such as sigmoid, hyperbolic tangent, or rectified linear unit (ReLU), is utilized to introduce nonlinearity to the model \cite{Goodfellow2016}. In this work, ReLU is used for all activation layers except for the output layer, as follows:
\begin{equation}
\label{eq:convof}
\phi_{g}^\zeta = {\rm max}(0,\phi_{g}^\zeta).
\end{equation}
Unlike sigmoid and hyperbolic tangent activation functions,  ReLU is robust to the vanishing gradient, thus, allowing deep models to learn faster and perform better \cite{Goodfellow2016}.

\textbf{Batch Normalization Layer:}
A batch normalization layer is added after the activation layer to avoid \textit{internal covariate shift}, which leads to an exponential increase in computation burden by requiring much lower learning rates \cite{SI2015}. Thus, the output of each activation layer is normalized by subtracting the batch mean and dividing by the batch standard deviation for each training mini-batch.

\textbf{Max-pooling Layer:}
After batch normalization, a max-pooling layer is utilized to summarize feature maps. Max-pooling is a sample-based discretization procedure that takes the feature map from the previous layer: $\phi_{g}^\zeta \in \mathbbm{R}^{ N_{\rm{in}} \times N_{\rm{in}}}$ and outputs a smaller matrix, denoted as $N_{\rm{out}} \times N_{\rm{out}}$. This is achieved by dividing the input matrix into $N_{\rm{out}}^2$ pooling regions $P_{i,j}$ and selecting the maximum value \cite{BG2014}:
\begin{equation}
\label{eq:pooling}
P_{i,j} \subset \{1,2,...,N_{\rm{in}}\}^2, \forall (i,j)\in\{1,2,...,N_{\rm{out}}\}^2.
\end{equation}
In our work, a $2 \times 2$ max-pooling is used, as shown in Fig. \ref{fig:layer} (b). Thus, $N_{\rm{in}}=2N_{\rm{out}}$ and $P_{i,j}=\{2i-1,2i\}\times \{2j-1,2j\}$. The functions of the max-pooling layer generalize the results from the convolutional-normalization operation and reduce the model complexity while alleviating overfitting.

\textbf{SPP Layer:}
In the proposed model, an SPP layer is adopted to replace the last max-pooling layer \cite{SPP}. Unlike the max-pooling layer, which performs a single pooling operation, the SPP layer maintains spatial information by pooling in local spatial bins, as shown in Fig. \ref{fig:layer}(c). This figure provides an exemplary 3-level SPP layer. Suppose the last convolutional layer has $r$ feature maps. In the first level, one bin is utilized to pool each feature map to become one value, thus, forming an $r$-dimensional vector. Then, four bins are leveraged to divide each feature map into 4 regions of equal size with a rectangular shape. The max-pooling strategy is applied to each region to form a $4\times r$-dimensional vector. In the final level, each feature map is pooled to have $16$ values, and form a $16\times r$-dimensional vector. In general, the outputs of the SPP are $r \cdot B$-dimensional vectors, where $B$ is the number of spatial bins, which is proportional to the MTF-graph size but is fixed. Hence, the output of the SPP layer is the fixed-dimensional vectors regardless of input size. In other words, after offline training, missing/bad data points can be directly removed during online testing. The remaining good-quality-data of arbitrary dimension can be assigned as the input of the proposed method.

\textbf{Fully-connected Layer:}
The last layer of the proposed method is a fully-connected layer, which integrates information across all locations in all the feature maps from the SPP layer. In this fully-connected layer, the softmax activation function is applied to calculate probabilities of the input being in a particular type. 

\begin{table}[tbp]
\caption{The structure of the SPP-aided CNN-based model.}
\centering
\renewcommand{\arraystretch}{0.85}
\begin{tabular}{cccc}
\hline\hline
Layout & Type & Output Shape & Model Complexity\\[1pt]
\hline
1/1 & Conv2D & (120,120,32) & 608\\[2pt]
1/2 & Activation & (120,120,32) & 0\\[2pt]
1/3 & Batch Norm & (120,120,32) & 128\\[2pt]
\hline
2/1 & Conv2D & (120,120,32) & 9k\\[2pt]
2/2 & Activation & (120,120,32) & 0\\[2pt]
2/3 & Batch Norm & (120,120,32) & 128\\[2pt]
2/4 & Max-pooling & (60,60,32) & 0\\[2pt]
\hline
3/1 & Conv2D & (60,60,64) & 18k\\[2pt]
3/2 & Activation & (60,60,64) & 0\\[2pt]
3/3 & Batch Norm & (60,60,64) & 256\\[2pt]
\hline
4/1 & Conv2D & (60,60,64) & 36k\\[2pt]
4/2 & Activation & (60,60,64) & 0\\[2pt]
4/3 & Batch Norm & (60,60,64) & 256\\[2pt]
4/4 & Max-pooling & (30,30,64) & 0\\[2pt]
4/5 & Dropout & (30,30,64) & 0\\[2pt]
\hline
5/1 & Conv2D & (30,30,128) & 73k\\[2pt]
5/2 & Activation & (30,30,128) & 0\\[2pt]
5/3 & Batch Norm & (30,30,128) & 512\\[2pt]
\hline
6/1 & Conv2D & (30,30,128) & 147k\\[2pt]
6/2 & Activation & (30,30,128) & 0\\[2pt]
6/3 & Batch Norm & (30,30,128) & 512\\[2pt]
6/4 & Max-pooling & (15,15,128) & 0\\[2pt]
6/5 & Dropout & (30,30,64) & 0\\[2pt]
6/6 & SPP & (1,2688) & 0\\[2pt]
\hline
7/1 & Fully-connected & (1,1,5) & 13k\\[2pt]
7/2 & Activation & (1,1,5) & 0\\[2pt]
\hline\hline
\end{tabular}
\label{table:1.2}
\end{table}

In the proposed SPP-aided CNN-based method, the adaptive moment estimation (Adam) algorithm is used to update the weight and bias variables \cite{adam}. Adam is a combination of gradient descent with momentum and root mean square propagation algorithms. Compared to backpropagation algorithms with constant learning rates (i.e., stochastic gradient descent), Adam computes individual adaptive learning rates for each parameter from estimates of first and second moments of the gradients \cite{adam}, which significantly increases the training speed. To calibrate the hyperparameters of the proposed method, we utilize the random search method to find the appropriate sets \cite{randomsearch}. Moreover, the dropout strategy is utilized in our model to further reduce the risk of overfitting.

\section{Numerical Results}\label{result}
To validate the effectiveness of the proposed event identification method, we test it on the recorded PMU datasets and the related event logs from interconnection B. This includes around 4800 events that consist of line outage, XFMR outage, frequency event, and oscillation event. Moreover, the same number of the 2-second analysis-window in normal conditions have been added in the event dataset. After data cleaning, the event dataset is randomly divided into two separate subsets for training (80$\%$ of the total data) and testing (20$\%$ of the total data). The case study is conducted on a standard PC with an Intel(R) Xeon(R) CPU running at 4.10GHZ and with 64.0GB of RAM and an Nvidia Geforce GTX 1080ti 11.0GB GPU. The average online computation time for performing the proposed method is around 1.4 ms. This implies that the method can be applied for real-time event identification, in accordance with the IEEE C37.118.2-2011 standard. 

\subsection{Performance of the Proposed Method}
\begin{figure}[tbp]
	\centering
	\includegraphics[width=3.2in]{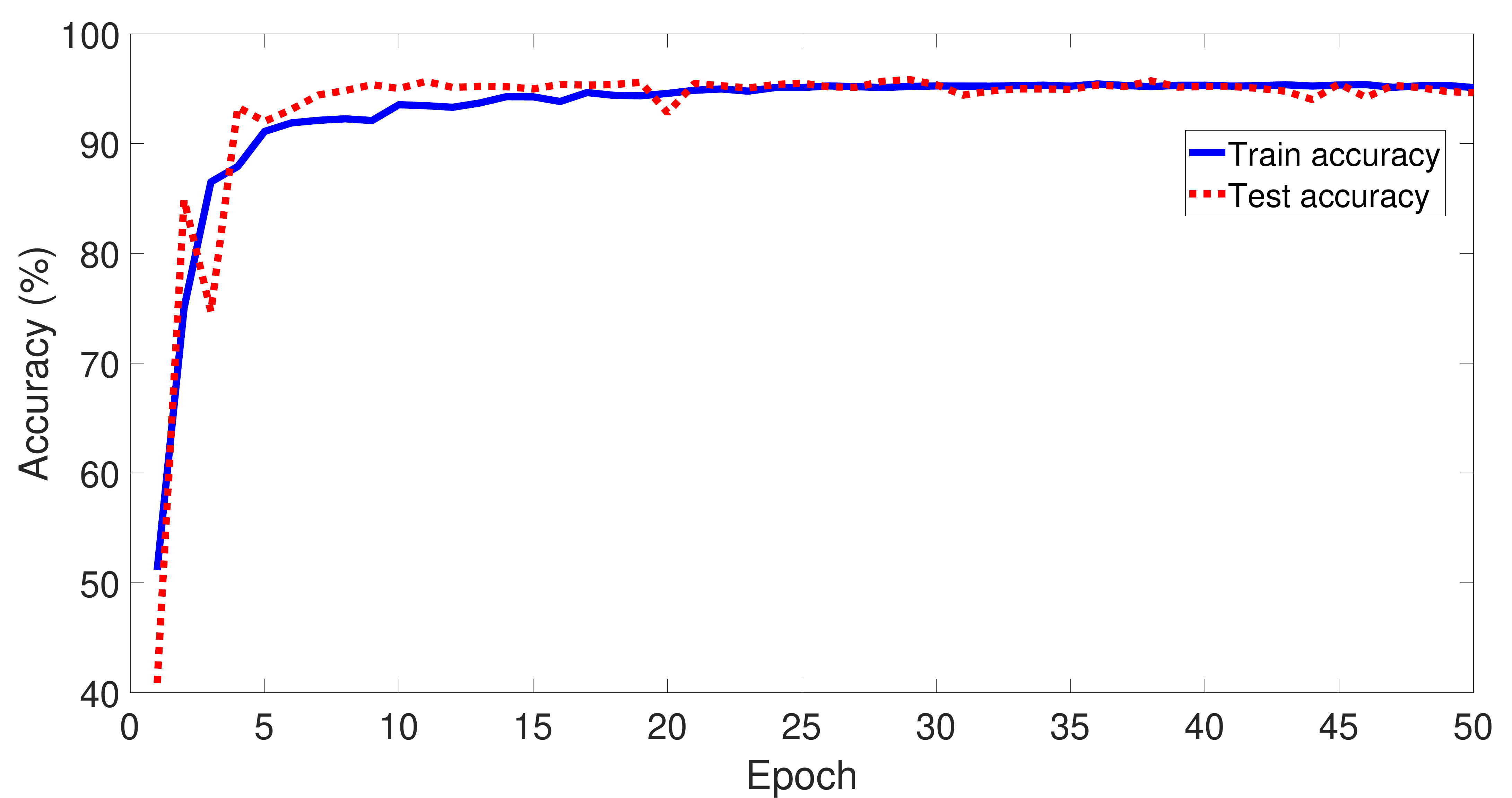}
	\caption{Training result for a SPP-aided CNN-based model.}
	\label{fig:train_test_acc}
	\vspace{-1em}
\end{figure}

\begin{figure}[tbp]
	\centering
	\includegraphics[width=3.2in]{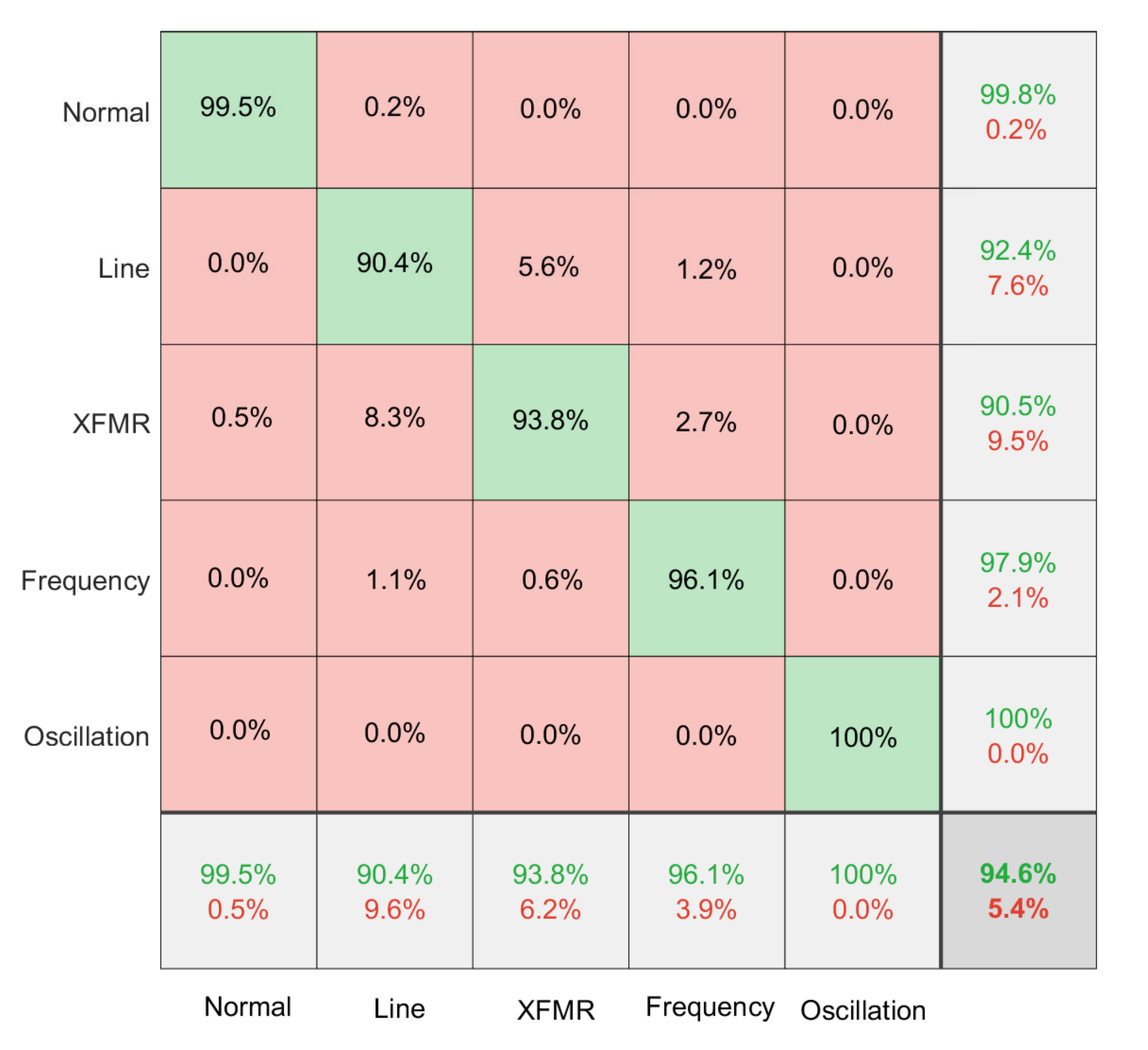}
	\caption{Confusion matrix for our PMU dataset using the proposed model.}
	\label{fig:confu}
	\vspace{-1em}
\end{figure}

The detailed structure of the proposed classifier is presented in Table \ref{table:1.2}. As can be seen, our model is a seven-layer architecture that includes multiple convolutional, activation, batch normalization, SPP, and fully-connected layers. Each row represents layers with specific layer type, the dimension of output feature map, and model complexity. Based on this structure, the training and testing performances of the proposed method are shown in Fig. \ref{fig:train_test_acc}. As demonstrated in this figure, the training and testing accuracy converge to around $95.1\%$ and $94.6\%$. The difference between the two values is small, which proves the generalization capability of the proposed model. Moreover, the performance of the proposed method is explained through the confusion matrices shown in Fig. \ref{fig:confu}. In this figure, the rows correspond to the predicted class and the columns correspond to the true class. The diagonal and off-diagonal cells correspond to events that are correctly and incorrectly classified, respectively. The value of each cell represents the accuracy of the specific event class. Several statistical scores are included: the precision and the recall rates are presented in the cells on the far right and the bottom of the figure, respectively. The cell in the bottom right of the figure is the overall accuracy. As seen in this figure, the worst-case precision and recall rates of the proposed method are around $90.5\%$ and $90.4\%$ for the XFRM outage and line outage classes, which still are acceptable values.

In practice, operators are interested in knowing a single system-level classification outcome rather than multiple PMU-level outcomes. Hence, we have obtained and tested the system-level results by collecting the classification outcomes of all PMUs: for a specific event, if more than 90$\%$ of PMU-level outcomes are positive, the event is identified at the system-level, using the proposed method. In this case, the system-level accuracy of our technique is around $91.07\%$.

\begin{figure}[tbp]
	\centering
	\includegraphics[width=3.2in]{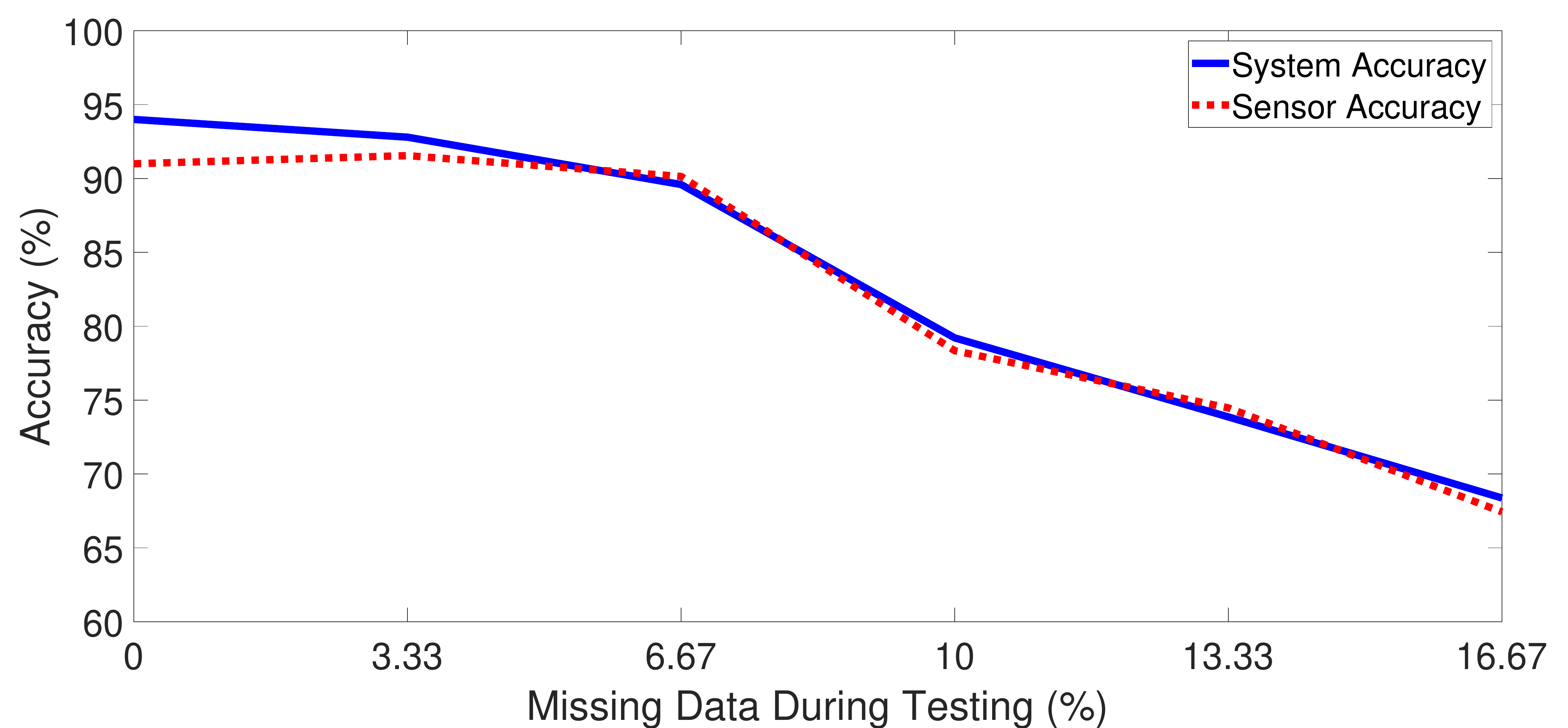}
	\caption{Sensitivity of event identification accuracy to the size of missing data.}
	\label{fig:sen}
\end{figure}

Moreover, to represent the sensitivity of the event identification accuracy to the size of missing data, we have shown the average performance of the proposed method under various sizes of consecutive missing data in Fig. \ref{fig:sen}. Note that the location of missing data is selected randomly and the results are obtained based on multiple Monte Carlo simulations. As is presented in the figure, the accuracy of our model decreases as the percentage of missing data increases due to the information loss. However, the proposed learning-based method still can achieve around $80\%$ accuracy when $10\%$ of data suffers from quality issue.

\subsection{Method Comparison}
\begin{figure}[tbp]
	\centering
	\includegraphics[width=3.2in]{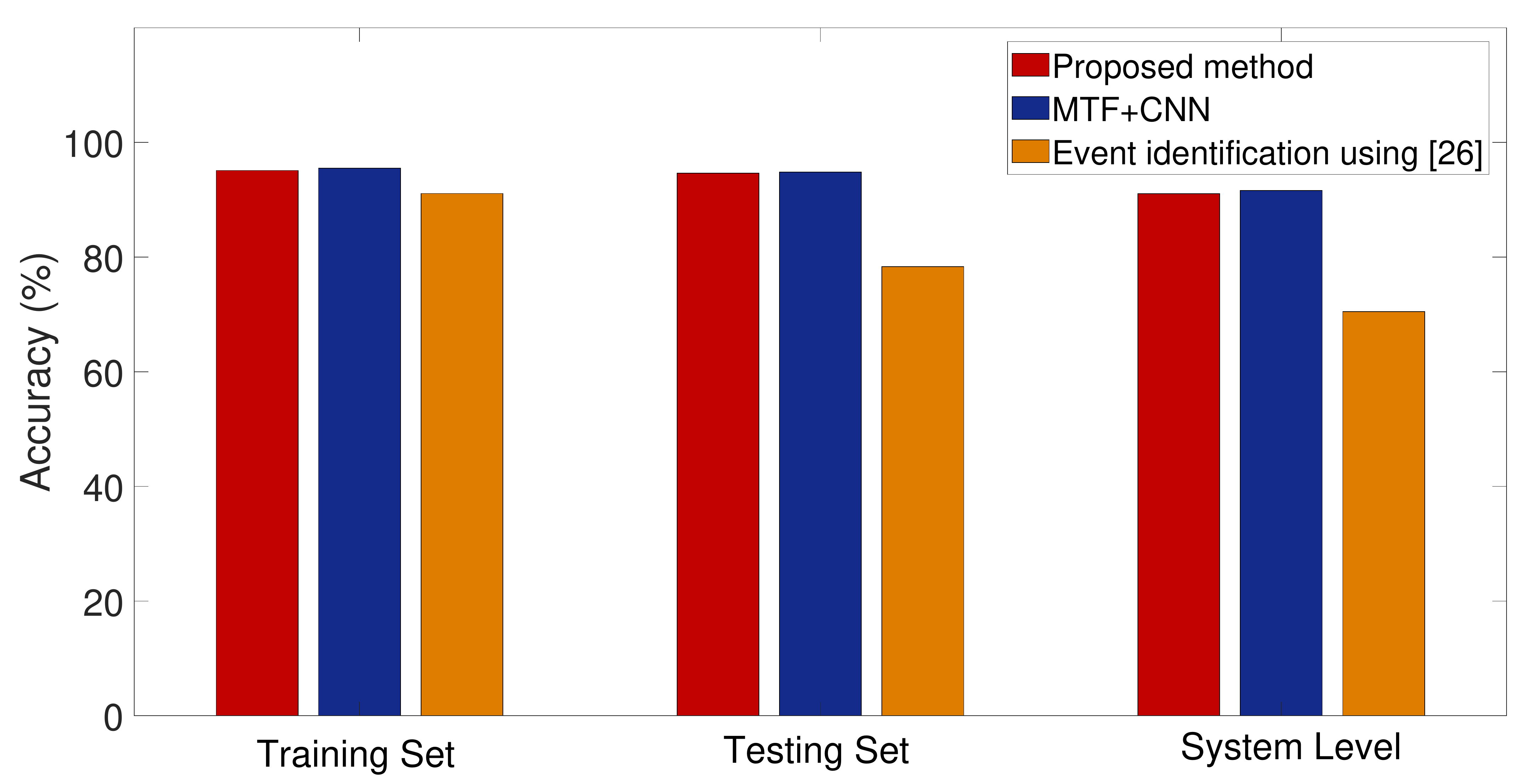}
	\caption{Comparison results with two methods.}
	\label{fig:compare}
\end{figure}

We have conducted numerical comparisons with two methods: a previous data-driven model \cite{compare} and the proposed method without SPP, to show that our method can address the data quality problem, as well as achieving good event identification accuracy. As is demonstrated in Fig. \ref{fig:compare}, the proposed method achieves similar accuracy with the model that only includes MTF and CNN. This indicates that the SPP strategy does not impact the identification performance; however, SPP is needed for resolving online data quality issues. Compared to the previous model \cite{compare}, our method shows a better accuracy for event identification using PMU data. 

\section{Conclusion}\label{conclusion}
In this paper, we have presented a novel two-stage learning-based method for real-time event identification to enhance the situational awareness of power systems using PMU data. Comparisons with previous methods show that our method achieves more accurate event identification outcomes. Moreover, this approach shows robustness against data quality problems in online testing, which highly improves the practical applicability in real-world systems. The proposed method is successfully validated on a large-scale PMU dataset and the related event logs.

\section*{Acknowledgment and Disclaimer}\label{ack}
This material is based upon work supported by the Department of Energy under Award Number DEOE0000910. This report was prepared as an account of work sponsored by an agency of the United States Government. Neither the United States Government nor any agency thereof, nor any of their employees, makes any warranty, express or implied, or assumes any legal liability or responsibility for the accuracy, completeness, or usefulness of any information, apparatus, product, or process disclosed, or represents that its use would not infringe privately owned rights.  Reference herein to any specific commercial product, process, or service by trade name, trademark, manufacturer, or otherwise does not necessarily constitute or imply its endorsement, recommendation, or favoring by the United States Government or any agency thereof. The views and opinions of authors expressed herein do not necessarily state or reflect those of the United States Government or any agency thereof.


\ifCLASSOPTIONcaptionsoff
  \newpage
\fi



\bibliographystyle{IEEEtran}
\bibliography{IEEEabrv,./bibtex/bib/IEEEexample}

\begin{thebibliography}{10}
\providecommand{\url}[1]{#1}
\csname url@samestyle\endcsname
\providecommand{\newblock}{\relax}
\providecommand{\bibinfo}[2]{#2}
\providecommand{\BIBentrySTDinterwordspacing}{\spaceskip=0pt\relax}
\providecommand{\BIBentryALTinterwordstretchfactor}{4}
\providecommand{\BIBentryALTinterwordspacing}{\spaceskip=\fontdimen2\font plus
\BIBentryALTinterwordstretchfactor\fontdimen3\font minus
  \fontdimen4\font\relax}
\providecommand{\BIBforeignlanguage}[2]{{%
\expandafter\ifx\csname l@#1\endcsname\relax
\typeout{** WARNING: IEEEtran.bst: No hyphenation pattern has been}%
\typeout{** loaded for the language `#1'. Using the pattern for}%
\typeout{** the default language instead.}%
\else
\language=\csname l@#1\endcsname
\fi
#2}}
\providecommand{\BIBdecl}{\relax}
\BIBdecl

\bibitem{2003bo}
D.~White, A.~Roschelle, P.~Peterson, D.~Schlissel, B.~Biewald, and
  W.~Steinhurst, ``The 2003 blackout: Solutions that won’t cost a fortune,''
  \emph{The Electricity Journal}, vol.~16, no.~9, pp. 43--53, 2003.

\bibitem{Ian}
W.~{Ju}, I.~{Dobson}, K.~{Martin}, K.~{Sun}, N.~{Nayak}, I.~{Singh},
  H.~{Saravia}, A.~{Faris}, L.~{Zhang}, and Y.~{Wang}, ``Real-time monitoring
  of area angles with synchrophasor measurements,'' \emph{arXiv preprint
  arXiv:2003.06476v1}, 2020.

\bibitem{DK2017}
D.~{Kim}, T.~Y. {Chun}, S.~{Yoon}, G.~{Lee}, and Y.~{Shin}, ``Wavelet-based
  event detection method using pmu data,'' \emph{IEEE Trans. on Smart Grid},
  vol.~8, no.~3, pp. 1154--1162, 2017.

\bibitem{MC2019}
M.~{Cui}, J.~{Wang}, J.~{Tan}, A.~R. {Florita}, and Y.~{Zhang}, ``A novel event
  detection method using pmu data with high precision,'' \emph{IEEE Trans.
  Power Systems}, vol.~34, no.~1, pp. 454--466, 2019.

\bibitem{YG2015}
Y.~Ge, A.~J. Flueck, D.~K. Kim, J.~B. Ahn, J.~D. Lee, and D.~Y. Kwon, ``Power
  system real-time event detection and associated data archival reduction based
  on synchrophasors,'' \emph{IEEE Trans. on Smart Grid}, vol.~6, no.~4, pp.
  2088--2097, 2015.

\bibitem{EP2008}
E.~Perez and J.~Barros, ``A proposal for on-line detection and classification
  of voltage events in power systems,'' \emph{IEEE Trans. Power Deli.},
  vol.~23, no.~4, pp. 2132--2138, Oct. 2008.

\bibitem{MB2016}
M.~{Biswal}, S.~M. {Brahma}, and H.~{Cao}, ``Supervisory protection and
  automated event diagnosis using pmu data,'' \emph{IEEE Trans. Power Deli.},
  vol.~31, no.~4, pp. 1855--1863, 2016.

\bibitem{SL2020}
S.~{Liu}, Y.~{Zhao}, Z.~{Lin}, Y.~{Liu}, Y.~{Ding}, L.~{Yang}, and S.~{Yi},
  ``Data-driven event detection of power systems based on unequal-interval
  reduction of pmu data and local outlier factor,'' \emph{IEEE Trans. Smart
  Grid}, vol.~11, no.~2, pp. 1630--1643, 2020.

\bibitem{JM2012}
J.~Ma, Y.~V. Makarov, R.~Diao, P.~V. Etingov, J.~E. Dagle, and E.~D. Tuglie,
  ``The characteristic ellipsoid methodology and its application in power
  systems,'' \emph{IEEE Trans. Power Systems}, vol.~4, no.~27, pp. 2206--2214,
  May 2012.

\bibitem{SB2017}
S.~{Brahma}, R.~{Kavasseri}, H.~{Cao}, N.~R. {Chaudhuri}, T.~{Alexopoulos}, and
  Y.~{Cui}, ``Real-time identification of dynamic events in power systems using
  pmu data, and potential applications—models, promises, and challenges,''
  \emph{{IEEE} Trans. Power Deli.}, vol.~32, no.~1, pp. 294--301, Feb. 2017.

\bibitem{LH2019}
H.~{Li}, Y.~{Weng}, E.~{Farantatos}, and M.~{Patel}, ``An unsupervised learning
  framework for event detection, type identification and localization using
  pmus without any historical labels,'' \emph{2019 IEEE Power Energy Society
  General Meeting (PESGM)}, pp. 1--5, 2019.

\bibitem{VC2007}
V.~Chandola, A.~Banerjee, and V.~Kumar, ``Anomaly detection: A survey,''
  \emph{Technical Report, University of Minnesota}, pp. 1--1, 2007.

\bibitem{JZ2019}
J.~{Zhao}, J.~{Tan}, L.~{Wu}, L.~{Zhan}, W.~{Yao}, and Y.~{Liu}, ``Impact of
  the measurement errors on synchrophasor-based wams applications,'' \emph{IEEE
  Access}, vol.~7, pp. 143\,960--143\,972, 2019.

\bibitem{OPD2012}
O.~P. {Dahal} and S.~M. {Brahma}, ``Preliminary work to classify the
  disturbance events recorded by phasor measurement units,'' \emph{2012 IEEE
  Power and Energy Society General Meeting}, pp. 1--8, 2012.

\bibitem{sat2006}
N.~Sokolova, Marinaand~Japkowicz and S.~Szpakowicz, \emph{Beyond Accuracy,
  F-Score and ROC: A Family of Discriminant Measures for Performance
  Evaluation}.\hskip 1em plus 0.5em minus 0.4em\relax Berlin, Heidelberg:
  Springer Berlin Heidelberg, 2006.

\bibitem{WZ2015}
Z.~Wang and T.~Oates, ``Encoding time series as images for visual inspection
  and classification using tiled convolutional neural networks,''
  \emph{Association for the Advancement of Artificial Intelligence}, pp.
  40--46, 2015.

\bibitem{CA2011}
A.~S. Campanharo, M.~I. Sirer, R.~D. Malmgren, F.~M. Ramos, and L.~A.~N.
  Amaral, ``Duality between time series and networks,'' \emph{PloS one},
  vol.~6, no.~8, pp. 40--46, 2011.

\bibitem{auto2017}
T.~V, C.~E, and L.~A, ``A deep convolutional auto-encoder with
  pooling-unpooling layers in caffe,'' \emph{arXiv preprint arXiv:1701.04949},
  2017.

\bibitem{Goodfellow2016}
I.~Goodfellow, Y.~Bengio, and A.~Courville, \emph{Deep Learning}.\hskip 1em
  plus 0.5em minus 0.4em\relax MIT Press, 2016,
  \url{http://www.deeplearningbook.org}.

\bibitem{SPP}
K.~{He}, X.~{Zhang}, S.~{Ren}, and J.~{Sun}, ``Spatial pyramid pooling in deep
  convolutional networks for visual recognition,'' \emph{IEEE Trans. on Pattern
  Analysis and Machine Intell.}, vol.~37, no.~9, pp. 1904--1916, 2015.

\bibitem{YQ2016}
Y.~{Qian}, M.~{Bi}, T.~{Tan}, and K.~{Yu}, ``Very deep convolutional neural
  networks for noise robust speech recognition,'' \emph{IEEE/ACM Trans. on
  Audio, Speech, and Language Processing}, vol.~24, no.~12, pp. 2263--2276,
  Dec. 2016.

\bibitem{SI2015}
S.~Ioffe and C.~Szegedy, ``Batch normalization: Accelerating deep network
  training by reducing internal covariate shift,'' \emph{arXiv preprint
  arXiv:1502.03167}, 2015.

\bibitem{BG2014}
B.~Graham, ``Fractional max-pooling,'' \emph{arXiv preprint arXiv:1412.6071},
  2014.

\bibitem{adam}
D.~P. Kingma and J.~Ba, ``Adam: A method for stochastic optimization,''
  \emph{arXiv preprint arXiv:1412.6980}, 2014.

\bibitem{randomsearch}
J.~Bergstra and Y.~Bengio, ``Random search for hyper-parameter optimization,''
  \emph{Journal of Machine Learning Research}, vol.~13, pp. 281--305, Feb.
  2012.

\bibitem{compare}
S.~Basumallik, R.~Ma, and S.~Eftekharnejad, ``Packet-data anomaly detection in
  pmu-based state estimator using convolutional neural network,''
  \emph{International Journal of Electrical Power \& Energy Systems}, vol. 107,
  pp. 690--702, 2019.

\end{thebibliography}

\end{document}